\documentclass[9pt,twocolumn,twoside]{opticajnl}
\journal{opticajournal} 

\setboolean{shortarticle}{true}

\usepackage{lineno}

\usepackage{chemformula} 
\usepackage{gensymb}
\usepackage{float}
\DeclareMathAlphabet\mathbfcal{OMS}{cmsy}{b}{n}

\title{Precision Doppler Shift Measurements with a Frequency Comb Calibrated Laser Heterodyne Radiometer}

\author[1,*]{Ryan K. Cole}
\author[1,2]{Connor Fredrick}
\author[3]{Newton H. Nguyen}
\author[1,2,4]{Scott A. Diddams}

\affil[1]{Time and Frequency Division, National Institute of Standards and Technology, Boulder CO, USA}
\affil[2]{Department of Physics, University of Colorado Boulder, Boulder, CO, USA}
\affil[3]{Division of Geological and Planetary Sciences, California Institute of Technology, Pasadena, CA, USA}
\affil[4]{Department of Electrical, Computer, and Energy Engineering, University of Colorado Boulder, Boulder, CO, USA}

\affil[*]{ryan.cole@nist.gov, ryan.cole@colorado.edu}

\begin{abstract}
We report precision atmospheric spectroscopy of \ch{CO2} using a laser heterodyne radiometer (LHR) calibrated with an optical frequency comb. Using the comb-calibrated LHR, we record spectra of atmospheric \ch{CO2} near 1572.33~nm with a spectral resolution of 200~MHz using sunlight as a light source. The measured \ch{CO2} spectra exhibit frequency shifts by approximately 11~MHz over the course of the five-hour measurement, and we show that these shifts are caused by Doppler effects due to wind along the spectrometer line of sight. The measured frequency shifts are in excellent agreement with an atmospheric model, and we show that our measurements track the wind-induced Doppler shifts with a relative frequency precision of 100~kHz ($\mathrm{15 \ cm \cdot s^{-1}}$), equivalent to a fractional precision of a few parts in 10$^{10}$. These results demonstrate that frequency-comb-calibrated LHR enables precision velocimetry that can be of use in applications ranging from climate science to astronomy. 
\end{abstract}

\setboolean{displaycopyright}{false} 

\begin{document}
\maketitle

Laser heterodyne radiometry (LHR) is a well known approach for spectroscopy of thermal light \cite{parvitte_infrared_2004}. In LHR, light from a continuous wave laser (the local oscillator, LO) is interfered with light from a thermal source, and the resulting heterodyne signal gives a measure of the power of the thermal light within a narrow frequency range around the LO laser. By tuning the LO laser frequency, a high-resolution optical spectrum (e.g. $R=\nu/\delta\nu \sim 10^6$) can be recorded within the scan range of the laser without the use of moving components or diffractive optics. 

Numerous past studies have demonstrated LHR with sunlight to record spectra of atmospheric trace gases \cite[e.g.]{bomse_precision_2020,deng_development_2021, hoffmann_thermal_2016, rodin_high_2014, tsai_atmospheric_2012, wilson_miniaturized_2014, weidmann_high-resolution_2009} or to study absorption transitions in the sun itself \cite{sappey_passive_2020, fredrick_thermal-light_2022}. Recently, Fredrick \textit{et al.} \cite{fredrick_thermal-light_2022} introduced LHR with a frequency comb calibration, bringing the absolute stability and traceability of the frequency comb to high-precision spectroscopy of solar absorption lines and transitions in a laboratory gas cell. Here, we extend this frequency-comb-calibrated LHR approach to atmospheric spectroscopy of greenhouse gases, and we show that the high spectral resolution and frequency precision of comb-calibrated LHR enables tracking of wind-induced Doppler shifts in the measured spectra with $\mathrm{cm\cdot s^{-1}}$ precision. 

While LHR is a well established technique for measuring mixing ratios of greenhouse gases and other atmospheric trace gases \cite[e.g.]{bomse_precision_2020,deng_development_2021, hoffmann_thermal_2016, rodin_high_2014, tsai_atmospheric_2012, wilson_miniaturized_2014, weidmann_high-resolution_2009}, several studies have also demonstrated that LHR is capable of atmospheric wind measurements through the Doppler shifts imparted by wind along the spectrometer line of sight \cite{rodin_vertical_2020,li_high-resolution_2023, goldstein_absolute_1991}. Atmospheric wind measurements are relevant in applications ranging from meteorology \cite{baker_lidar-measured_2014} to climate and greenhouse gas monitoring \cite{world_meteorological_organization_wmo_19th_2018}. For example, wind drives the transport of atmospheric greenhouse gases, and when combined with coincident mixing ratio data, wind speed measurements provide an important constraint in our understanding of the spatiotemporal gradients of greenhouse gases and other atmospheric trace gases \cite{world_meteorological_organization_wmo_19th_2018}. To this end, expanding the remote sensing capabilities of LHR to include atmospheric wind measurements could provide valuable climate and meteorological data to complement measurements based on more established techniques (e.g. Doppler radar, lidar or microwave radiometry). More broadly, extending the capabilities of LHR for Doppler velocimetry could expand the utility of LHR in applications beyond climate and meteorology such as precision Doppler spectroscopy of astronomical sources \cite{fischer_state_2016} or passive tracking of thermal objects.

\begin{figure*}[h!]
\centering
\includegraphics[width=1\textwidth]{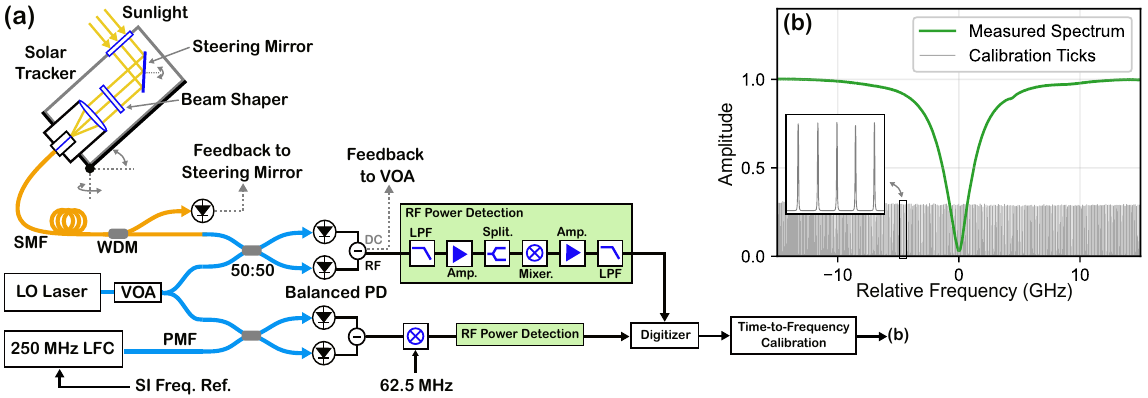}
\caption{Panel (a): Schematic of the frequency-comb-calibrated LHR approach (see text for description). Panel (b): Five-hour-averaged spectrum of the \ch{CO2} R16 transition near 1572.33 nm. The inset shows the calibration tick marks (spaced by $f_r/2 = 125\ \textrm{MHz}$) measured by interfering the LO laser with the frequency comb.}
\label{figure1}
\end{figure*}

Spectroscopic wind measurements pose a demanding challenge for the spectrometers used to make the measurement. For example, resolving a Doppler shift due to 1~$\mathrm{m\cdot s^{-1}}$ line of sight motion requires a spectrometer with fractional frequency precision ($\delta f / f$) better than $10^{-9}$. Recent LHR-based wind measurements have addressed this challenge with a frequency calibration based on an etalon \cite{rodin_high_2014} or Mach-Zehnder interferometer \cite{li_high-resolution_2023} in combination with a reference gas cell that is used to determine the line center of the target transition in the rest frame. With this approach, these studies have reported vertically-resolved measurements of absolute wind speeds with precision at the meter-per-second level \cite{rodin_vertical_2020, li_high-resolution_2023}. Here, we address the challenge of frequency stability by calibrating our LHR system with a laser frequency comb to enable spectroscopy of sunlight with the stability and frequency accuracy of a frequency comb. Using this approach, we track wind-induced Doppler shifts in measured spectra with precision better than 100~kHz ($\sim$15 $\mathrm{cm\cdot s^{-1}}$).

Figure \ref{figure1} shows a schematic of our frequency-comb-calibrated LHR approach. This apparatus has been described in detail in Ref. \cite{fredrick_thermal-light_2022}, and here we list only the salient details. We couple solar light into single-mode fiber using a solar-tracking telescope. The telescope consists of a commercial solar tracker (EKO STR-22G) and piezo-actuated steering mirror that directs solar light onto a fiber collimator. The steering mirror provides secondary pointing corrections to account for small deviations in the solar tracking. The steering mirror pointing is locked to the bright center of the solar disk with feedback based on the fiber-coupled solar power measured after splitting the solar light in a 1310/1550~nm WDM. A refractive beam shaper placed between the steering mirror and the fiber collimator uniformly integrates light from the solar disk by transforming the Gaussian fiber mode to a flat-top profile in the far field \cite{fredrick_thermal-light_2022}. 

Fiber-coupled solar light is combined with light from a DFB diode laser (the LO) that is temperature-tuned over the target absorption transition. The solar and LO light are combined in a polarization-maintaining 50:50 fiber coupler and interfered on a balanced photodetector (Thorlabs PDB465A). The radio frequency~(RF) output of the photodetector is sent to an RF power detection circuit (described below), while the DC monitor output is used to feed back to a variable optical attenuator~(VOA) that stabilizes the LO laser power and mitigates signal distortion due to variations in laser power during each scan. 

The RF power detection circuit is the same as described in Ref. \cite{fredrick_thermal-light_2022}. Briefly, the heterodyne output of the photodetector is amplified and passed through a low pass filter (LPF) that sets the spectral resolution of the measurement as twice the filter cutoff frequency. The filtered signal is split in a power splitter and passed to both inputs of a double-balanced mixer. The mixer output is terminated into 50 ohms, and the resulting DC voltage is proportional to the heterodyne signal power. The DC signal is passed through a preamplifier and additional low pass filter before being digitized on an oscilloscope. 

In a second channel, the LO light is simultaneously interfered with light from a stabilized, $f_r = 250 \ \mathrm{MHz}$ Er:fiber frequency comb. The heterodyne signal between the LO and comb is recorded on a balanced detector and mixed with a synthesized 62.5 MHz tone that doubles the density of the frequency calibration points \cite{fredrick_thermal-light_2022,jennings_frequency_2017}. The RF power detection circuit is the same as described above, but uses a lower filter cutoff (2 MHz) that limits the heterodyne signal to a narrow range around each comb mode. The output of this process is a series of calibration "ticks" that occur whenever the scanning LO laser coincides with a comb mode. 

The output of the comb-calibrated LHR system described above is a DC signal proportional to the spectrum of the solar light and a simultaneously recorded series of frequency calibration ticks. We determine the frequency axis of the measured spectra by fitting each calibration tick with a Gaussian profile to determine its centroid. Using the resulting calibration points as well as the known frequency spacing between each point ($f_R/2$), we construct a time-to-frequency transfer function that transforms the temporal axis of the measurement to the comb-referenced frequency grid. 
The frequency comb used for this comb calibration is referenced to a NIST-calibrated hydrogen maser and provides a SI-traceable frequency calibration grid with relative uncertainty of a few parts in $10^{13}$ or better. Accounting for the relative uncertainty in the maser comb reference and the time-to-frequency calibration process, we estimate the relative frequency uncertainty of the comb calibration to be $\sim$70~kHz for a single measurement (10~s), averaging to $\sim$5~kHz at one hour. At that level, line center determination is limited by noise in the measured spectra.

\begin{figure*}[t!]
\centering
\includegraphics[width = \textwidth]{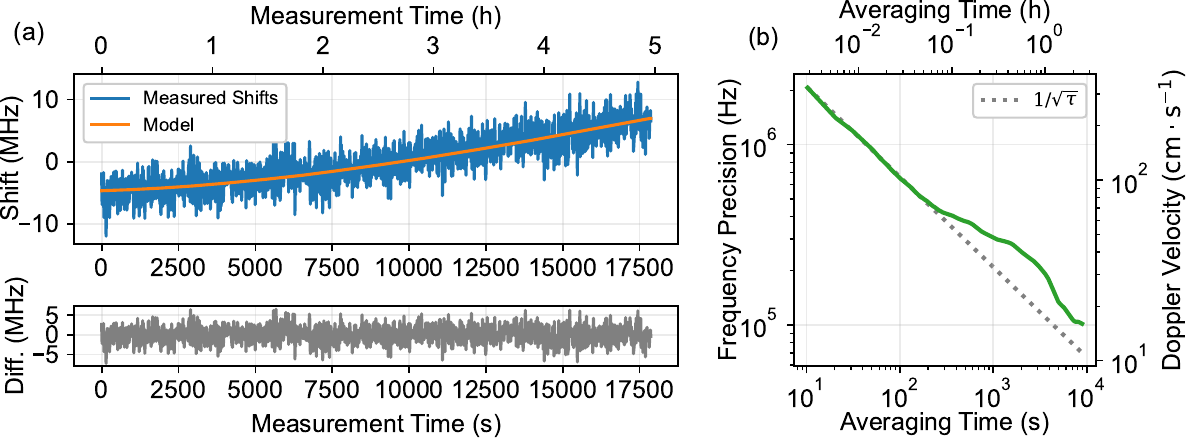}
\caption{Panel (a): Comparison between the measured frequency shifts and the shifts predicted using the atmospheric model. The lower panel shows the differences between measured and modeled shifts. Measurement times are specified relative 10:10 a.m. local time (UTC-6). Panel (b): Allan deviation of the difference between the measured and modeled frequency shifts.}
\label{results}
\end{figure*}

Using the approach outlined above, we recorded spectra for atmospheric \ch{CO2} in Boulder, CO, USA on October 12, 2022. The measurement targeted the R16 transition of the $30012\leftarrow00001$ \ch{CO2} band near 1572.33 nm, which has been the subject of past remote sensing missions \cite{abshire_pulsed_2010} and advanced spectroscopic characterization \cite{long_air-broadened_2011}. Figure \ref{figure1}(b) shows the measured \ch{CO2} spectrum after averaging for nearly five hours. The spectrum was recorded using a low pass filter bandwidth of 100 MHz, which results in a measurement spectral resolution of 200 MHz. The effective averaging time (2~ms) of the final low pass filter in the RF detection chain (see Figure \ref{figure1}) yields $\sim$30 independent samples per 200~MHz resolution element. Each spectrum was recorded in a 10~s scan spanning a $\sim$30 GHz optical window. The signal-to-noise ratio (SNR) for each 10~s scan is $\sim$50. Owing to the high stability of the comb calibration, long-term averaging of the measured spectra allows the SNR to grow with $\sqrt{\tau}$, exceeding 2000 after averaging for the full measurement period. 


We assess the relative frequency precision of the measured spectra by comparing each measurement to a template generated from the five-hour-averaged spectrum. The observed shift in each spectrum is taken as the frequency shift that maximizes the cross correlation between the spectrum and the template. In this sense, this approach determines frequency shifts relative to the spectrum averaged over the full measurement period. For the \ch{CO2} spectra measured on October 12, the frequency shifts indicate a progressive blue shift by $\sim$11 MHz over the course of the five-hour measurement. Figure \ref{results}(a) shows the measured frequency shifts along with a comparison to our model for the expected frequency shifts due to wind-induced Doppler effects along the LHR line of sight.

To model the effect of wind on the measured spectra, atmospheric temperature, pressure, and three-dimensional wind fields are obtained from the European Center for Medium-Range Weather Forecast ERA5 reanalysis data \cite{hersbach_era5_nodate} for Boulder, CO, USA. The ERA5 data has a temporal resolution of one hour, and we linearly interpolate the data to estimate the atmospheric conditions for each measured LHR spectrum. We split the atmosphere into 50 altitude bins, and simulate the \ch{CO2} R16 transition in each layer using the HITRAN2020 database with temperature-dependent line shape parameters for the speed-dependent Nelkin-Ghatak profile (SDNGP) \cite{gordon_hitran2020_2021}. We assume a uniform \ch{CO2} mixing ratio of 400 ppm, which after integrating over the 50 atmospheric layers produces a simulated line shape that is in qualitative agreement with measured spectra.

\begin{figure}[t!]
\centering
\includegraphics[width = \linewidth]{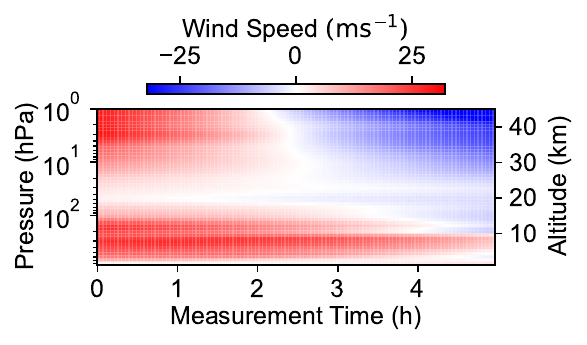}
\caption{Line-of-sight wind speeds for October 12, 2022 determined using the ERA5 wind fields. Times are specified relative to the start of data collection (10:10 a.m. local time, UTC-6).}
\label{windspeed}
\end{figure}

Our model accounts for wind-induced Doppler effects by applying a frequency shift to the simulated spectrum in each atmospheric layer. The wind speed along the LHR line of sight (and therefore the Doppler shift) is determined as 
    \begin{equation}
    \mathcal{W}^{LOS}_{i} = \mathbfcal{W}_i\cdot \mathbf{\hat{k}}
    \label{eq1}
    \end{equation}
where $\mathcal{W}^{LOS}_{i}$ is the line-of-sight wind speed in layer $i$, $\mathbf{\hat{k}}$ is the normalized LHR pointing vector, and $\mathbfcal{W}_i$ is the wind velocity vector in terms of eastward ($\mathbf{\hat{u}}$), northward ($\mathbf{\hat{v}}$), and downward ($\mathbf{\hat{w}}$) components. The pointing vector is specified in the ($u,v,w$) coordinate system as 
    \begin{equation}
    \mathbf{k} = \left ( sin \theta_z sin \alpha \right ) \mathbf{\hat{u}} + \left ( sin \theta_z cos \alpha \right ) \mathbf{\hat{v}} + \left ( cos \theta_z \right ) \mathbf{\hat{w}}
    \label{eq2}
    \end{equation}
where $\theta_z$ is the solar zenith angle and $\alpha$ is the azimuth angle. We determine the solar position angles for each measured LHR spectrum using a Python wrapper for NREL's Solar Position Algorithm \cite{f_holmgren_pvlib_2018,reda_solar_2004}. Figure \ref{windspeed} shows the line of sight wind speeds calculated using this method for the data on October 12. 

After simulating spectra at times corresponding to each measured LHR spectrum, we determine the wind-induced Doppler shifts using the same cross correlation approach described above. In this case, the wind-induced shifts are determined relative to a template generated by averaging the simulated spectra over the five-hour measurement period.  Figure \ref{results}(a) shows the wind-induced shifts calculated using our model, which are in excellent agreement with the measured shifts over the full duration of the measurement. Figure \ref{results}(b) shows the Allan deviation of the difference between the measured and calculated shifts. For a single spectrum (10 s), we track the line center with a precision of $\sim$2~MHz (3~$\mathrm{m \cdot s^{-1}}$) and the precision improves to approximately $\sim$100 kHz (15 $\mathrm{cm \cdot s^{-1}}$) after 2.5 hours of averaging. Relative to the $\sim$2.5 GHz line width of the measured transition, this frequency precision splits the line by a factor of 25,000.

In evaluating the results shown in Figure \ref{results}, it is also important to consider how atmospheric variability (e.g. changes in temperature and pressure) could influence the observed line shift. To assess the strength of these effects relative to wind-induced Doppler shifts, we reran the atmospheric model while including variability in the atmospheric pressure and temperature but neglecting Doppler effects. For the data on October 12, surface pressure increased from approximately 834 to 836 hPa over the course of data collection based on measurements from a co-located weather station. Using our model, we estimate that an increase in atmospheric pressure at this level could affect a frequency shift of approximately 80 kHz over the course of the five-hour measurement with a sign opposite that of the wind-induced shifts. Similarly, we use the ERA5 data to estimate changes in atmospheric temperature, and we find that temperature variability induces shifts by $\sim$40 kHz. In both cases, the pressure- and temperature-induced shifts are small relative to wind-induced Doppler shifts. 

Furthermore, although our analysis has involved only relative frequency shifts, it is interesting to consider the use of comb-calibrated LHR to measure absolute shifts (and thus absolute wind speeds). Past LHR-based wind measurements have determined vertically-resolved, absolute wind speeds using inversion methods that rigorously fit the measured spectra with an atmospheric model \cite{li_high-resolution_2023, rodin_vertical_2020}. This approach represents a significant increase in complexity when compared to relative shift measurements, which are only concerned with deviations from the average. The relative shifts shown in Figure \ref{results} depend only on the stability of the spectrometer, and a measurement of absolute shifts would depend on additional factors such as the accuracy of the atmospheric and spectroscopic data used to fit the measured spectra. Nonetheless, comb-calibrated LHR may still provide valuable benefits for absolute wind measurements by leveraging the stability and absolute frequency accuracy of the comb calibration to reduce instrumental uncertainties and enable precision tracking of absolute Doppler shifts over long time scales. Future studies could explore how these benefits impact absolute wind measurements when combined with a more advanced retrieval procedure.


In conclusion, we demonstrate high-precision spectroscopy of atmospheric \ch{CO2} through the unique combination of a laser heterodyne radiometer and an optical frequency comb. We show that our measurements track wind-induced Doppler shifts in the measured \ch{CO2} spectra with a precision of $\sim$100 kHz ($\mathrm{15 \ cm \cdot s^{-1}}$), equivalent to a fractional frequency precision of a few parts in $10^{10}$. These results demonstrate the potential of frequency-comb-calibrated LHR as an approach for precision atmospheric spectroscopy and Doppler metrology. LHR has a long heritage as a technique for remote sensing of greenhouse gas mixing ratios, and future efforts could seek to combine these established capabilities with precision Doppler wind measurements. Such efforts could significantly expand the capabilities of LHR as a climate monitoring tool and provide valuable data to constrain emissions estimates and greenhouse gas transport.

More broadly, our results validate comb-calibrated LHR as a tool for precision Doppler velocimetry that could be of use in applications beyond climate monitoring. Such applications may include passive tracking of thermal objects or precision radial velocity measurements of astronomical sources, including characterizing the impact of telluric absorption on those measurements \cite{fischer_state_2016, figueira_evaluating_2010}. In the latter application, achieving precision Doppler spectroscopy at the $\mathrm{cm \cdot s^{-1}}$ levels represents an ongoing challenge in the fields of solar and exoplanet science that could be explored in future studies using comb-calibrated LHR.

\begin{backmatter}

\bmsection{Funding} This work was supported by the NIST IMS program, NIST financial assistance award 70NANB18H006, and the NASA Astrophysics Division. R.C. acknowledges support from the National Academies NRC Research Associateship Program. 

\bmsection{Acknowledgments} The authors thank Eugene Tsao and David Plusquellic for valuable comments and discussions. This work is a contribution of NIST and is not subject to copyright in the United States. Mention of specific products or trade names is for technical and scientific information and does not constitute an endorsement by NIST.

\bmsection{Disclosures} The authors declare no conflicts of interest.

\end{backmatter}

\bibliography{NIST}

\end{document}